\documentclass[journal,twoside]{IEEEtran}
\usepackage{amsmath,amsfonts}
\usepackage{algorithmic}
\usepackage{array}
\usepackage{bm}
\usepackage[caption=false,font=normalsize,labelfont=sf,textfont=sf]{subfig}
\usepackage{textcomp}
\usepackage{stfloats}
\usepackage{verbatim}
\usepackage{graphicx}
\usepackage{cite} 
\usepackage{ifthen}  
\usepackage{multicol} 
\usepackage{dsfont} 
\usepackage{amssymb}
\usepackage{color} 
\usepackage{soul} 
\usepackage{booktabs}
\usepackage{ctable} 
\usepackage{balance}
\usepackage{tabularx} 
\usepackage{url}
\usepackage{hyperref}
\makeatletter
\newcommand{\thickhline}{%
    \noalign {\hrule height 1pt}%
}
\newcolumntype{'}{!{\vrule width 1pt}}
\makeatother 
\def\BibTeX{{\rm B\kern-.05em{\sc i\kern-.025em b}\kern-.08em
    T\kern-.1667em\lower.7ex\hbox{E}\kern-.125emX}}
\begin{document}
\title{Performance Evaluation of Neuromorphic Hardware for Onboard Satellite Communication Applications}
\author{Eva~Lagunas,~\IEEEmembership{Senior Member,~IEEE,} Flor~G.~Ortiz,~\IEEEmembership{Member,~IEEE,} Geoffrey~Eappen,~\IEEEmembership{Member,~IEEE,} Saed~Daoud,~\IEEEmembership{Member,~IEEE,} Wallace~A.~Martins,~\IEEEmembership{Senior Member~IEEE}, Jorge~Querol,~\IEEEmembership{Member,~IEEE,} Symeon~Chatzinotas,~\IEEEmembership{Fellow~IEEE,} Nicolas~Skatchkovsky,~\IEEEmembership{Member,~IEEE,} Bipin~Rajendran,~\IEEEmembership{Senior Member~IEEE}, and Osvaldo~Simeone,~\IEEEmembership{Fellow~IEEE.}}
\markboth{IEEE Communications Magazine}{E. Lagunas \MakeLowercase{\textit{(et al.)}: Performance Evaluation of Neuromorphic Hardware for Onboard Satellite Communication Applications}}
\maketitle
\begin{abstract}
Spiking neural networks (SNNs) implemented on neuromorphic processors (NPs) can enhance the energy efficiency of deployments of  artificial intelligence (AI) for specific workloads. As such, NP represents an interesting opportunity for implementing AI tasks on board power-limited satellite communication spacecraft. In this article, we disseminate the findings of a recently completed study which targeted the comparison in terms of performance and power-consumption of different satellite communication use cases implemented on standard AI accelerators and on NPs. In particular, the article describes three prominent use cases, namely payload resource optimization, onboard interference detection and classification, and dynamic receive beamforming; and compare the  performance of conventional convolutional neural networks (CNNs) implemented on Xilinx's VCK5000 Versal development card and SNNs on Intel's neuromorphic chip Loihi 2.
\end{abstract}
\section{Introduction}
\label{sec:intro}
\IEEEPARstart{A}{rtificial} intelligence (AI) has been identified as an essential ingredient of the next generation of wireless communications \cite{9145564}. AI opens up exciting opportunities, including autonomous network reconfiguration to changing environments, improved system performance, and enhanced customer experience. Given the benefits of AI in the terrestrial wireless environment,  the satellite communication (SatCom) research community has also started exploring  AI for various  applications. 

Traditionally, SatCom payloads have been built based on hardware components generally conceived for dedicated tasks. Software-defined radio (SDR) based technology has proven to be beneficial to the satellite industry \cite{9275613}. By replacing hardware components with software, SDR allows a more flexible satellite payload, offering system reconfiguration in real time.
 
Existing works in the open  literature have analyzed  the adoption of machine learning (ML) algorithms for SDR-based SatCom use cases  at the level of concepts and simulations\cite{Vazquez2021MachineOperations,fontanesi2023artificial}. While testing in these works relies on software models, their successful transition to real deployments must hinge on their efficient implementation on chipsets that are able to perform the task within reasonable accuracy, computation time, and  power budget. This is particularly relevant for onboard AI implementation, where the AI chipset is to be integrated within the satellite payload. 

Neuromorphic processors (NPs) represent a potentially revolutionary solution for efficiently deploying AI models targeting specific workloads. NPs implement energy-efficient spiking neural networks (SNNs) that mimic the operation of the human brain.  Given the stringent onboard power limitations of satellite platforms, NPs represent a major opportunity to unlock the potential benefits of AI and ML solutions for SatCom systems, thanks to their energy efficiency. Power must be generated onboard, typically exploiting the sunlight using solar panels. However, the size of such panels dictate the size of the satellite and determines the launching cost, being one of the most critical design aspects of space missions.

Industry and academia are currently working on the development of specific AI processors that  provide better energy and latency performance for computationally intensive AI algorithms. Some of these AI processors are already available on the market, while many are still in testing and design phase. In this work, we first discuss the methodology for shortlisting the most promising SatCom use cases, and provide a description of the three selected ones, which are: (1) Resource optimization in flexible satellite payloads; (2) Onboard interference detection and classification; and (3) Dynamic digital beamforming for fast-moving users. Subsequently, we discuss and analyze the representative SatCom use cases using two promising alternative commercial options, namely Xilinx's VCK5000 Versal development card for classical AI deployments and Intel's Loihi 2 chipset for neuromorphic platforms. The first is an acceleration chip from the Versal family already commercialized by Xilinx, which was specifically designed to implement high throughput AI inference and signal processing computing performance. The latter corresponds to the second generation of Intel's neuromorphic research chips. We quantify gains in terms of computational time, power consumption, reliability, and processor footprint, as well as performance on the specific communications-related task. The research activity presented in this manuscript was carried out in collaboration with the European Space Agency (ESA) in the context of the ESA NeuroSat project \cite{NeuroSat}. 

This article is organized as follows. First, we describe the selected use cases in Section~\ref{sec:Appl}. Next, we discuss and compare the two selected AI chipsets in Section~\ref{sec:chips}, and discuss the data encoding methodologies for neuromophic implementation in Section~\ref{sec:SNN}. Finally, we present the performance comparisons in Section~\ref{sec:Perf}, followed by the conclusion in Section~\ref{sec:conclu}. 

\section{Selected SatCom Use Cases}
\label{sec:Appl}
In recent years, AI has been applied for a plethora of SatCom use cases \cite{Vazquez2021MachineOperations,fontanesi2023artificial}. Such widespread interest is a natural response to the emergence of a new technology, and it highlights the need for a clear identification of the most  practically relevant use cases. 

\begin{figure}[h!]
  \centering
  \centerline{\includegraphics[scale=0.5]{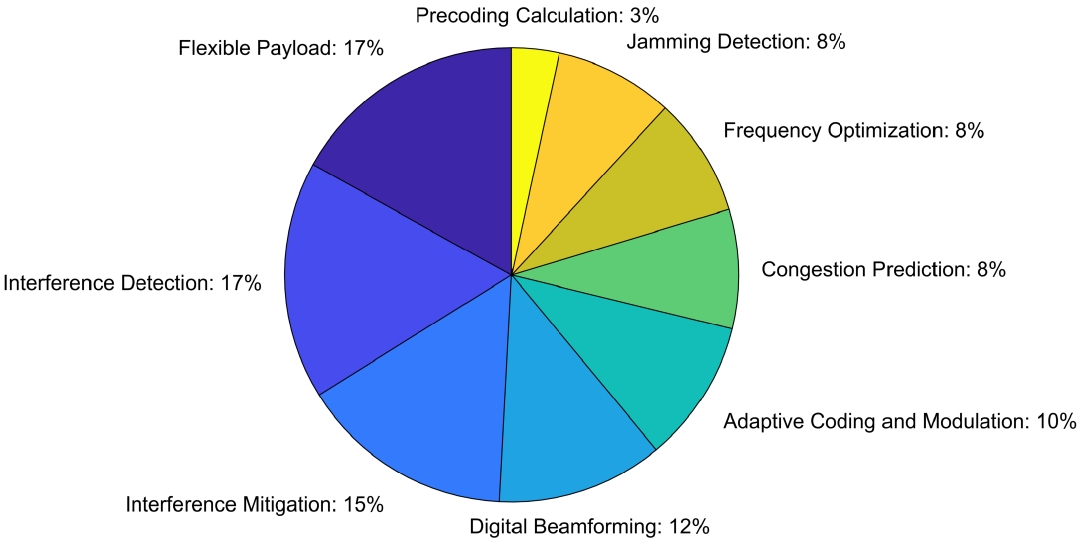}}
\caption[Text excluding the matrix]{Outcome of use case relevance for AI application from  survey conducted on industry and academia representatives.  \cite{NeuroSat} }
\label{WorkshopNeuroSat_fig}
\end{figure}

In the context of the ESA project {NeuroSat} \cite{NeuroSat}, a workshop including both commercial AI-chip vendors and SatCom service providers, as well as leading European research institutions, was held in July 2022 with the aim of collecting feedback from experts regarding the most promising and feasible scenarios and use cases for the implementation of learning-based methods. Details on the workshop are available on the project website \cite{NeuroSat}. The participants were asked to vote for the most promising scenario based on their expertise. The outcomes of such voting are detailed in Fig.~\ref{WorkshopNeuroSat_fig}. Notably, SatCom service providers identified use cases such as flexible payload operations and interference detection and mitigation as critical, given their current needs and the recent advances in network reconfiguration and the increasing spectrum congestion. In contrast, they tended to give less importance to well-studied problems for which they already have efficient solutions based on classical optimization such as precoding matrix calculation. For NP experts, having the possibility of a good temporal encoding for the corresponding AI-model inputs was found to be crucial to obtain gains via NP deployments. 

Based on the preferences of stakeholders and considering the potential gains that AI could offer, the following use cases were chosen for further evaluation:
\begin{enumerate}
  \item resource optimization in flexible satellite payloads.
  \item onboard interference detection and classification.
  \item dynamic digital beamforming for fast-moving users.
\end{enumerate}

In the following subsections, we present the details of the shortlisted use cases.

\subsection{Resource Optimization in Flexible Satellite Payloads}
\label{sec:rrm}

Dynamically adapting the radio resource allocation to match the spatiotemporal on-ground traffic demand variations is a challenging task for the new generation of software-defined satellite payloads \cite{9237970}. Power and bandwidth assignment per beam are the two elements typically considered as degrees of freedom to achieve a good match between the users' on-ground demand and the actual capacity offered by the communication satellite. 

The input of the resource optimization is given by the on-ground traffic demand. Such input considers a geographical map of the service area, divided into a grid of latitude and longitude points, forming a matrix. This matrix represents the traffic demand at each point on the map. Each element $r_{i,j}$ in this matrix represents the required data traffic (in Mbps) at a specific geographic location indexed by  $i$ and $j$. Each location within the satellite service area provides information about its data traffic needs, while areas outside the service zone are irrelevant and are thus assigned zero values.

The objective is to determine the optimal combination of power and bandwidth settings for each satellite beam within our service area. We create ``payload configurations'' consisting of power and bandwidth pairs for each beam. A particular payload configuration selected for a specific time $\tau$ can be denoted as $\left[ (P_{\tau}^1,{\rm W}_{\tau}^{1}), (P_{\tau}^2,{\rm W}_{\tau}^{2}), \ldots, (P_{\tau}^K,{\rm W}_{\tau}^{K}) \right]$, where $k\in\{1,\ldots, K\}$ denotes the index of satellite beams. The selected power and bandwidth belong to a discrete set, i.e., $P_{\tau}^k \in \mathcal{P}_{\tau}$ and ${\rm W}_{\tau}^{k} \in \mathcal{W}_{\tau}$. However, it is crucial to recognize that not all combinations might be feasible due to practical constraints on total power and bandwidth. 

The proposed approach is a classification problem, where a CNN takes as input the geographical traffic demand distribution and selects the appropriate payload configuration. To minimize complexity in the number of outputs (i.e., possible payload configurations), instead of considering all possible combinations of power and bandwidth, we make a simplification by: (1) Considering only the combinations of power and bandwidth options that comply with the system constraints in terms of maximum power and bandwidth; and (2) Pre-training the system and removing the payload configurations from the candidate pool that are rarely selected for accommodating the traffic demand patterns expected by the system. Once the output possibilities are narrowed down, supervised learning comes into play to match traffic demand with the optimal payload configuration.

The optimization task for flexible satellite payload is summarized in Table~\ref{rrm_tab}.

\begin{table}[htb]
\centering
\caption{Satellite Payload Configuration Optimization}
\label{rrm_tab}
  \begin{tabular}{| c | p{5cm} |  }
	\thickhline
\textbf{Input} &  On-ground traffic demand in Mbps per latitude and longitude grid point. \\	\thickhline
\textbf{Output} &  A selection from a limited set of power-per-beam and bandwidth-per-beam configurations.\\	\thickhline
\textbf{Objective} &  Minimize the difference between on-ground traffic demand and offered capacity.\\	\thickhline
\textbf{Type of Problem} &  Classification, i.e., matching each input with one of the potential outputs.\\	\thickhline
  \end{tabular}
\end{table}

\subsection{Onboard Interference Detection and Classification}
\label{sec:idc}

The intended or unintended interference experienced in space, either generated by ground transmitters or by other SatCom systems, represents one of the major problems of the satellite industry, which is being aggravated by the trends of launching more and more in-orbit satellite systems, particularly the popular low-Earth orbit (LEO) constellations. Interference events may occur due to intentional jamming, or more likely due to equipment and/or human errors (e.g.,  antenna's misalignment and cross-polarization effects). For a satellite operator, the first step is to detect the interference event and characterize the received interference. In our approach, the interference detection and classification problem is reduced to a  classification problem with the following output classes:
\begin{itemize}
  \item The received signal is free of interference (class $0$).
  \item The received signal is interfered at subband $\mathcal{F}_i$, $i\in\{1,\ldots,L\}$ (classes $1,2,\ldots,L$).
\end{itemize}


The onboard interference detection task is summarized in Table~\ref{idc_tab}.

\begin{table}[htb]
\centering
\caption{Onboard interference Detection and Classification}
\label{idc_tab}
  \begin{tabular}{| c | p{5cm} |  }
	\thickhline
\textbf{Input} &  Frequency and/or time domain representation of the received signal. \\	\thickhline
\textbf{Output} &  The frequency band of the detected interference (out from a limited set of options).\\	\thickhline
\textbf{Objective} &  Detect the interference and classify the spectrum of interest.\\	\thickhline
\textbf{Type of Problem} &  Classification, i.e., matching each input with one of the potential output.\\	\thickhline
  \end{tabular}
\end{table}

\subsection{Dynamic Digital Beamforming for Fast-Moving Users}
\label{sec:beamf}
Receiver beamforming in the user uplink can significantly improve the received signal quality. However, it might be challenging to properly steer the beams toward users that move at  high speeds (e.g., aircraft), especially when it comes to non-geostationary orbits. For such high-speed users, the location information received from the gateway might be substantially outdated in scenarios with small spot-beams and, therefore, render beam-pointing errors. In this context, the problem boils down to selecting the optimum beamforming coefficients that allow to point the beam toward the fast-moving user. To address this problem, we use a least absolute shrinkage and selection operator (LASSO) regression to obtain the targeted beamforming vector, where the sparsity term is applied to the real and imaginary parts of the beamforming vector in an attempt to maximize the zero components and, thus, switch off as many radio frequency (RF) chains as possible. The beamforming task is summarized in Table~\ref{beamf_tab}.

\begin{table}[htb]
\centering
\caption{Dynamic digital beamforming for fast-moving users}
\label{beamf_tab}
  \begin{tabular}{| c | p{5cm} |  }
	\thickhline
\textbf{Input} &  Received signal at the satellite composed of a single uplink from a specific fast-moving aircrafts. \\	\thickhline
\textbf{Output} &  Onboard beamforming weights, which could in principle assume any complex value.\\	\thickhline
\textbf{Objective} &  A design for receive digital beamforming which can be implemented turning off as many RF chains as possible.\\	\thickhline
\textbf{Type of Problem} & LASSO regression optimization.\\	\thickhline
  \end{tabular}
\end{table}

\section{Selected AI Chipsets: VCK5000 vs. Loihi 2}
\label{sec:chips}
Although the field of AI has evolved rapidly in the past decades, computing processors on which AI algorithms run have not developed at the same pace, which may limit or delay their anticipated benefits. Existing computers are predominantly based on the classical von Neumann architecture, where computation and memory are implemented as separate elements connected via a common data bus, resulting in significant overheads, both in terms of latency and energy for shuttling data back and forth. This bottleneck is critical for AI workloads that involve constant  fetching of big amounts of data, impacting  power consumption and peak performance \cite{8887553}. As a response, the community has been intensively investigating AI hardware accelerators, i.e., dedicated and customized processors for AI-specific tasks.

A detailed overview of commercial off-the-shelf (COTS) AI-capable chipsets was presented in \cite{aerospace10020101}. Based on that, and on the current availability and lead times, we selected the Xilinx VCK5000 for the evaluation of  the machine learning models described in this article.

The Xilinx Versal VCK5000 is an ML-based development card built on Xilinx FPGAs and adaptive compute acceleration platforms. It is designed for high-efficiency AI acceleration with optimized deep learning processor unit cores for solving problems in 5G and beyond communication, signal processing, radar, and satellite-based applications. In the present scenario, VCK5000 is a fully developed AI chip capable of running CNNs, recurrent neural networks, and natural language processing-based models for cloud and edge applications. The training of a neural network model on VCK5000 employs three primary steps, which are quantization, compilation, and deployment. The insights obtained from training the CNN model on Versal VCK5000 is that the VCK5000 is a domain-specific architecture. The AI developer on VCK5000 should be specific about the deep learning processor unit, as it is related to the employed AI models. The training of CNN models on VCK5000 has high computational accuracy. This results in higher energy consumption as a trade-off. 

Recently, the concept of NP has gathered significant attention as an alternative architecture imitating the biological brains by operating in an event-driven fashion. NPs may therefore represent a suitable approach to unlocking the potential benefits of AI. While different companies are leading research efforts on the potential commercialization of NPs, commercial products are yet to be available at the moment of writing this paper. To carry out our study, we joined the Intel Neuromorphic Research Community (INRC), which provides access to the latest NP technology for research purposes. In particular, INRC provided access to the Loihi 2 chipset, which is a research neuromorphic chip that uses asynchronous spiking neurons to implement fine-grained, event-driven, adaptive, self-modifying, parallel computations. 
Loihi 2, the next-generation iteration of Loihi, serves as a follow-up to the original neuromorphic research test chip, Loihi. Within Loihi 2, multiple neuromorphic cores, each housing numerous artificial neurons, are interconnected, and they receive spikes from other parts of the network. When these received spikes accumulate over a specific time interval and surpass a predetermined threshold, the respective core initiates its own spike transmission to connected neurons. Previous spikes strengthen existing neuronal connections, whereas subsequent spikes inhibit these connections, gradually reducing connectivity until all activity ceases. Loihi 2 is constructed using the Intel 4 process and boasts a total of 1 million artificial neurons per chip and 120 million synapses per chip. In addition to the 128 neuromorphic cores, the chip incorporates 6 processor cores \cite{davies2021taking}.


\section{Data Encoding for Neuromorphic Implementation}
\label{sec:SNN}  


While an SNN can theoretically receive data in the form of an analog input current,
NPs can typically only handle data in the form of binary streams of spikes as inputs. Consequently, the natural signals we consider throughout the aforementioned   SatCom use cases  need to be encoded into binary spikes for processing via Intel's Loihi 2 chip.

The general encoding procedure is as follows. Considering a general feature matrix $\bm{x} \in \mathbb{R}^{n \times m}$, we first obtain an  $nm \times 1$ vector $[x_{1, 1} \dots x_{n,m}]^{\rm T}$, before performing the encoding process into a spiking signal $\bm{X} \in \{0, 1\}^{nm \times T}$, with $T$ being the number of encoding time-steps.

\begin{figure}[h!]
  \centering
  \centerline{\includegraphics[scale=0.45]{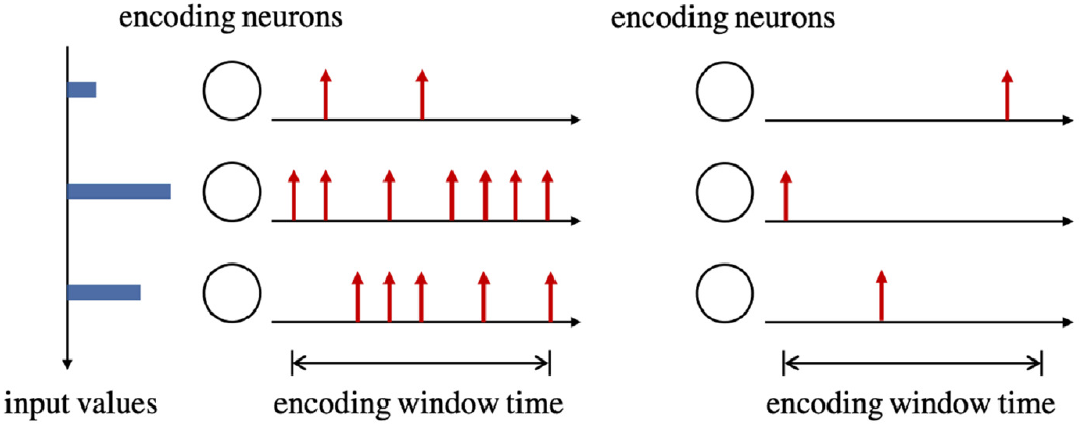}}
\caption[Text excluding the matrix]{Illustration of encoding schemes: rate coding (left) and temporal coding (right).}
\label{RateEncoding_fig}
\end{figure}

Among the different neural coding schemes \cite{10021878},  we focus on two of the most predominant ones: rate coding and temporal coding, both illustrated in Fig.~\ref{RateEncoding_fig} and briefly explained below:
\paragraph{rate coding} Rate coding utilizes spiking rates to represent information, i.e., larger real-valued input generates a large number of spikes within a fixed encoding window time.
\paragraph{temporal coding} Temporal coding utilizes precise spike arrival times to encode information. 

We now describe the details of signal encoding for the selected use cases.  

\subsection{Resource Optimization in Flexible Satellite Payloads}

The inputs to be encoded in the resource optimization use case described in Section~\ref{sec:rrm} correspond to the on-ground traffic demands in latitude and longitude, at different times of the day. Since the temporal aspect is intrinsically related to the input signal, we propose to use temporal coding for the on-ground traffic demand of each geographical location. Therefore, we end up with one time series for each geographical position. More precisely, we consider a temporal model based on leaky integrate-and-fire (LIF) neurons, whereby a spike is emitted when the ``voltage'' of the neuron (i.e., the traffic demand) crosses a predefined threshold.

\subsection{Onboard Interference Detection and Classification}

The use case scenario of the interference detection employs discrete Fourier transform (DFT), implemented via a fast Fourier transform (FFT) algorithm, on the input samples. The magnitude of the FFT signal can be encoded either with rate or temporal coding without problem. However, the preprocessing of the temporal signal to compute its spectral components via an auxiliary processor may incur a significant latency in a neuromorphic system. Recent works have demonstrated that resonate-and-fire (R\&F) neurons can compute the short-term Fourier transform (STFT) directly in the spiking domain \cite{9605018}. An R\&F neuron is an oscillatory system that operates by maintaining a complex-valued internal variable that spikes when its real part crosses a predetermined threshold and its argument is zero. For comparison purposes, we checked the computation of the STFT on an auxiliary processor and followed by temporal coding.

\subsection{Dynamic Digital Beamforming for Fast-Moving Users}

The LASSO beamforming solution does not involve a learning task but rather an optimization procedure. In this case, we selected the spiking locally competitive algorithm (S-LCA) to solve the problem, and we tested it using Intel’s Lava simulator. 




\section{Performance Comparison}
\label{sec:Perf}



\begin{figure}[htb]
  \centering
  \centerline{\includegraphics[scale=0.45]{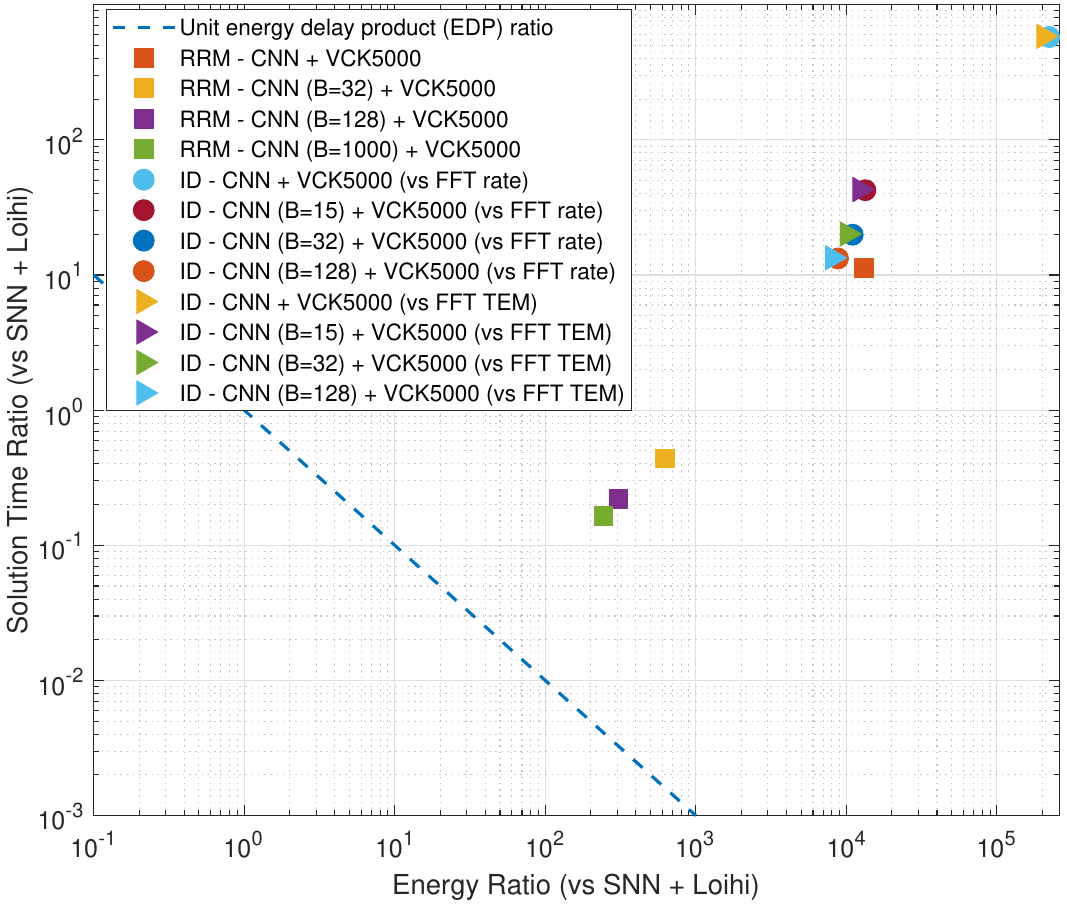}}
\caption[Text excluding the matrix]{Comparison in terms of energy ratio and delay ratio comparing the two chipsets (Xilinx's VCK5000 and Intel's Loihi 2) for different applications.}
\label{VCK_Loihi_fig}
\end{figure}

The graph in Fig.~\ref{VCK_Loihi_fig} displays the results for the first two  use cases, namely the resource optimization for flexible payloads (denoted by RRM) and the interference detection and classification (denoted by ID). Fig.~\ref{VCK_Loihi_fig} offers a comparison in a single plot of the two chipsets that have been studied, i.e., Xilinx's VCK5000 and Intel's Loihi 2, in terms of energy consumption and time taken to converge to a solution (delay).


To make sense of the graph, we follow  \cite{Mike2021IEEEProc} and adopt the energy-to-delay ratio (EDP) as a benchmark. This ratio serves as a reference point and is interpreted as follows:
\begin{itemize}
    \item \textbf{Points above the EDP:} Any data points located further to the right of the EDP line represent instances where Intel's Loihi 2 outperforms the Xilinx's VCK5000 in the sense that  Loihi 2 is more efficient in terms of power and time.
    \item \textbf{Points below the EDP:} Conversely, data points in the lower-left portion of the graph indicate scenarios where the reference architecture Xilinx's VCK5000 perform better than Loihi 2. These are situations in which the reference architectures are more efficient.
\end{itemize}

 Fig.~\ref{VCK_Loihi_fig} shows results for different batch sizes, denoted by $B$, i.e., the number of input samples given to the chip. Below we summarize the conclusions that can be extracted from Fig.~\ref{VCK_Loihi_fig}:
\begin{itemize}
\item  \textbf{Superiority of SNN and Loihi 2:} The results highlight that, in all cases, Intel's Loihi 2 performs better than the CNN implemented in Xilinx's VCK5000. However, it is worth noting that, as the batch size increases, the advantage of Loihi became less pronounced as the performance points move closer to the EDP line, although it still outperforms the Xilinx's VCK5000 implementation.

\item  \textbf{Interference detection as a promising use case:} Among the considered scenarios, it seems that the interference detection and classification benefited the most from the implementation on Intel's Loihi chipset. Even though the time ratio remained generally higher than one, the energy savings achieved with Loihi were significant. In some cases, the energy ratio reached values as high as $10^5$, which is remarkable.

\item \textbf{SNN encoding:} Fig.~\ref{VCK_Loihi_fig} compares the impact of FFT Rate and FFT TEM encoding for the ID scenario. Interestingly, the type of coding used did not have a significant impact on this comparison.

\item \textbf{RRM's Performance:} While RRM did not achieve energy savings as pronounced as in the ID scenario, it consistently achieved an energy ratio exceeding $10^2$ and always boasted a time ratio less than 1, highlighting its efficiency.
\end{itemize}

Regarding the digital beamforming performance for fast-moving users, we compared the conventional LASSO solution provided by CVX \cite{cvx} running on Matlab with the solution of S-LCA on Intel’s Lava simulator. Firstly, it is worth highlighting that the proposed beamforming formulation yielded sparse beamforming vectors, with both solutions being able to turn off up to 60\% of the RF chains without compromising the resulting beampatterns. Regarding performance comparisons between the two solutions, both generated satisfactory beampatterns with the main lobe pointing toward the aircraft. For a numerical comparison, the beamformer's average output power was considered as key performance indicator to assess the beamforming capabilities to mitigate the effects of noise and interference while focusing on the desired signal direction. In this context, the S-LCA solution was able to reach lower levels of beamformer's average output power, around 19\% below the value reached by the CVX solution, but with a much higher spreading of values, around 4 times higher than the CVX solution, when comparing the lower and upper quartiles of beamformer's average output power.


\subsection{Remarks about  the results obtained with Intel's Loihi 2}
The results presented in this article were conducted with Loihi 2 but using the remote access from Intel’s research cloud. Although Intel offers the possibility of shipping physical chips to INRC partners premises, at the moment of developing these results the primary access to Loihi 2 was through the Intel's neuromorphic research cloud. Obviously, the remote access introduced some additional limitations as it was not possible to control for concurrent usage by other users, which could lead to delays and increased power consumption. Additionally, the specific interface used by Intel on their cloud was not disclosed, potentially resulting in differences when conducting measurements with future releases of Loihi 2. Furthermore, due to the runtime limitation of 20~minutes for jobs on Intel’s cloud, on-chip training was not possible, restricting the exploration of this important capability of Loihi.

The cloud interface plays a key role, as it impacts the transfer of spiking signals to the chipset. High input size may span long per-step processing time. For example, in the flexible payload use case, the execution time per example increased from approximately 5 ms to 100 ms when the input size went from 252 neurons to 299 neurons.



\section{Conclusion}
\label{sec:conclu}
While we enter the era of AI, it becomes evident that energy consumption is a limiting factor when training and implementing neural networks with significant number of neurons. This issue becomes particularly relevant for nonterrestrial communication devices, such as satellite payloads, which are in need of more efficient hardware components in order to benefit from the potential of AI techniques.

Neuromorphic processors, such as Intel's Loihi 2, have shown to be more efficient when processing individual data samples and are, therefore, a better fit for use cases where real world data arrives to the chip and it needs to be processed right away. In this article, we verified this hypothesis using real standard processor solutions, such as the Xilinx's VCK5000 and the Intel's Loihi 2 chipset. 


\section*{Acknowledgments}
This work has been supported by the European Space Agency (ESA) funded under Contract No. 4000137378/22/UK/ND - The Application of Neuromorphic Processors to Satcom Applications. Please note that the views of the authors of this paper do not necessarily reflect the views of ESA. Furthermore, this work was partially supported by the Luxembourg National Research Fund (FNR) under the project SmartSpace (C21/IS/16193290).

\balance 
\bibliographystyle{IEEEbib}
\bibliography{refs_neurosat}
\section{Biography Section}
\vspace{11pt}
\begin{IEEEbiographynophoto}{Eva Lagunas}
received the M.Sc. and Ph.D. degrees in telecommunications engineering from the Polytechnic University of Catalonia (UPC), Barcelona, Spain, in 2010 and 2014, respectively. She has held positions at UPC, Centre Tecnologic de Telecomunicacions de Catalunya (CTTC), University of Pisa, Italy; and the Center for Advanced Communications (CAC), Villanova University, PA, USA. In 2014, she joined the Interdisciplinary Centre for Security, Reliability and Trust (SnT), University of Luxembourg, where she currently holds a Research Scientist position. Her research interests include terrestrial and satellite system optimization, spectrum sharing, resource management and machine learning.
\end{IEEEbiographynophoto}
\vspace{11pt}
\begin{IEEEbiographynophoto}{Flor G. Ortiz} received her B.S. degree in telecommunications engineering and her M.S. degree in electrical engineering-telecommunications from the Universidad Nacional Aut\'onoma de M\'exico (UNAM), Mexico City, Mexico, in 2015 and 2016, respectively. Flor obtained her Ph.D. degree in Telecommunication Engineering (September 2021) at the Universidad Politecnica de Madrid (UPM), Madrid, Spain. During her PhD, she performed a research period at the University of Bologna, in Bologna, Italy. She started a close collaboration between UPM and the University of Bologna, opening a new research line for both groups on applying Machine Learning for radio resource management. She is joined as a Research Associate at the Interdisciplinary centre for Security, Reliability, and Trust (SnT) at University of Luxembourg. Her research interests are focused on implementing cutting-edge Machine Learning techniques including Continual Learning and and Neuromorphic Computing for operations in Satellite Communications systems.
\end{IEEEbiographynophoto}
\vspace{11pt}
\begin{IEEEbiographynophoto}{Geoffrey Eappen}
(Member, IEEE), received the Ph.D. degree from the Vellore Institute of Technology (VIT), Vellore,
and Brunel University London. He was a Senior
Research Fellow with the Council of Scientific and
Industrial Research India, Department of Wireless
Communication (SENSE), VIT. He is currently
working as a Rsearch Associate with the Interdisciplinary Centre for Security, Reliability and Trust (SnT), University of Luxembourg. His research interest include cognitive
radio network, spectrum sensing methodologies,
metaheuristic optimization schemes, artificial neural
network, 5G beamforming, and satellite communi-
cation. He is the recipient of the U.K. Commonwealth Fellowship.
\end{IEEEbiographynophoto}
\begin{IEEEbiographynophoto}{Saed Daoud}
Saed Daoud received his M.Sc. degree from Jordan University of Science and Technology, Jordan in 2010, and 
his PhD degree from Concordia University, Montreal, Canada in 2015, both in Electrical Engineering with 
specialization in Wireless Communication. He was a postdoctoral researcher at Ecole Polytechnique de 
Montreal, and currently a research associate with the Interdisciplinary Centre for Security, Reliability and Trust 
(SnT), University of Luxembourg. His research interests broadly cover physical-layer transceiver design and 
performance evaluation for underwater acoustic (UWA), terrestrial and satellite communication systems, for 
technologies such as OFDM, SC-FDE, cognitive radio, spread spectrum, MIMO, and precoding.
\end{IEEEbiographynophoto}
\begin{IEEEbiographynophoto}{Wallace A. Martins}
received his D.Sc. degree in Electrical Engineering from the Federal University of Rio de Janeiro (UFRJ), Brazil, in 2011. He is currently a Full Professor with ISAE-SUPAERO, Université de Toulouse, France. His research interests encompass digital signal processing and telecommunications, focusing on future wireless and satellite networks.
\end{IEEEbiographynophoto}
\begin{IEEEbiographynophoto}{Jorge Querol}
Jorge Querol (S’15–M’18) was born in Forcall, Castelló, Spain, in 1987. He received the B.Sc. (+5) degree in telecommunication engineering, the M.Sc. degree in electronics engineering, the M.Sc. degree in photonics, and the Ph.D. degree (Cum Laude) in signal processing and communications from the Universitat Politècnica de Catalunya - BarcelonaTech (UPC), Barcelona, Spain, in 2011, 2012, 2013 and 2018 respectively. His Ph.D. thesis was devoted to the development of novel anti-jamming and counter-interference systems for Global Navigation Satellite Systems (GNSS), GNSS-Reflectometry, and microwave radiometry. One of his outstanding achievements was the development of a real-time standalone pre-correlation mitigation system for GNSS, named FENIX, in a customized Software Defined Radio (SDR) platform. FENIX was patented, licensed and commercialized by MITIC Solutions, a UPC spin-off company. 
Since 2018, he is with the SIGCOM research group of the Interdisciplinary Centre for Security, Reliability, and Trust (SnT) of the University of Luxembourg, Luxembourg and head of the 6GSPACE Laboratory. He is involved in several ESA and Luxembourgish national research projects dealing with signal processing and satellite communications. His research interests include SDR, real-time signal processing, satellite communications, 5G non-terrestrial networks, satellite navigation, and remote sensing. He received the best academic record award of the year in Electronics Engineering at UPC in 2012, the first prize of the European Satellite Navigation Competition (ESNC) Barcelona Challenge from the European GNSS Agency (GSA) in 2015, the best innovative project of the Market Assessment Program (MAP) of EADA business school in 2016, the award Isabel P. Trabal from Fundació Caixa d’Enginyers for its quality research during his Ph.D. in 2017, and the best Ph.D. thesis award in remote sensing in Spain from the IEEE Geoscience and Remote Sensing (GRSS) Spanish Chapter in 2019. 

\end{IEEEbiographynophoto}
\begin{IEEEbiographynophoto}{Symeon Chatzinotas} (MEng, MSc, PhD, FIEEE) is currently Full Professor / Chief Scientist I and Head of the research group SIGCOM in the Interdisciplinary Centre for Security, Reliability and Trust, University of Luxembourg. In parallel, he is an Adjunct Professor in the Department of Electronic Systems, Norwegian University of Science and Technology and a Collaborating Scholar of the Institute of Informatics \& Telecommunications, National Center for Scientific Research “Demokritos”. In the past, he has lectured as Visiting Professor at the University of Parma, Italy and contributed in numerous R\&D projects for the Institute of Telematics and Informatics, Center of Research and Technology Hellas and Mobile Communications Research Group, Center of Communication Systems Research, University of Surrey. He has received the M.Eng. in Telecommunications from Aristotle University of Thessaloniki, Greece and the M.Sc. and Ph.D. in Electronic Engineering from University of Surrey, UK in 2003, 2006 and 2009 respectively. He has authored more than 800 technical papers in refereed international journals, conferences and scientific books and has received numerous awards and recognitions, including the IEEE Fellowship and an IEEE Distinguished Contributions Award. He is currently in the editorial board of the IEEE Transactions on Communications, IEEE Open Journal of Vehicular Technology and the International Journal of Satellite Communications and Networking.
\end{IEEEbiographynophoto}
\begin{IEEEbiographynophoto}{Nicolas Skatchkovsky}~received his Ph.D. in Electrical Engineering from King's College London in 2022, under the supervision of Prof. Osvaldo Simeone. Before that, he received his M.Eng (with Honours) and B.Eng in Electrical Engineering from CentraleSup\'elec, Gif-sur-Yvette, France. He was a post-doctoral research associate at King's College London from 2022 to 2023, where he worked on Bayesian learning and applications of neuromorphic learning to space communications. He is now a post-doctoral research associate at the Francis Crick Institute, working on the modelling of attention in mamallian brains using machine learning techniques. His current research interests include applications of AI to neurosciences under an information-theoretic perspective.
\end{IEEEbiographynophoto}
\begin{IEEEbiographynophoto}{Bipin Rajendran} is a Professor of Intelligent Computing Systems in the Department of Engineering, King's College London, where he co-directs the Centre for Intelligent Information Processing (CIIPS). He received a B. Tech degree from I.I.T. Kharagpur in 2000, and M.S. and Ph.D. degrees in Electrical Engineering from Stanford University in 2003 and 2006, respectively. He was a Master Inventor and Research Staff Member at IBM T. J. Watson Research Center in New York during 2006-'12 and has held faculty positions in India and the US.

His research focuses on building algorithms, devices, and systems for intelligent computing systems. He has co-authored over 95 papers in peer-reviewed journals and conferences, one monograph, one edited book, and 59 issued U.S. patents. He is a recipient of the IBM Faculty Award (2019), IBM Research Division Award (2012), and IBM Technical Accomplishment Award (2010). He was elected a senior member of the US National Academy of Inventors in 2019.

His research has been supported by the Engineering and Physical Sciences Research Council (EPSRC), the US National Science Foundation (NSF), the European Commission, the European Space Agency, Semiconductor Research Corporation as well as Intel, IBM, and Cisco. In 2022, he was awarded an Open Fellowship of the EPSRC.

\end{IEEEbiographynophoto}
\begin{IEEEbiographynophoto}{Osvaldo Simeone}
is a Professor of Information Engineering. He co-directs the Centre for Intelligent Information Processing Systems within the Department of Engineering of King's College London, where he also runs the King's Communications, Learning and Information Processing lab. He received an M.Sc. degree (with honors) and a Ph.D. degree in information engineering from Politecnico di Milano, Milan, Italy, in 2001 and 2005, respectively. From 2006 to 2017, he was a faculty member of the Electrical and Computer Engineering (ECE) Department at New Jersey Institute of Technology (NJIT), where he was affiliated with the Center for Wireless Information Processing (CWiP). His research interests include information theory, machine learning, wireless communications, neuromorphic computing, and quantum machine learning. Dr Simeone is a co-recipient of the 2022 IEEE Communications Society Outstanding Paper Award, the 2021 IEEE Vehicular Technology Society Jack Neubauer Memorial Award, the 2019 IEEE Communication Society Best Tutorial Paper Award, the 2018 IEEE Signal Processing Best Paper Award, the 2017 JCN Best Paper Award, the 2015 IEEE Communication Society Best Tutorial Paper Award and of the Best Paper Awards of IEEE SPAWC 2007 and IEEE WRECOM 2007. He was awarded an Open Fellowship by the EPSRC in 2022 and a Consolidator grant by the European Research Council (ERC) in 2016. His research has been also supported by the U.S. National Science Foundation, the European Commission, the European Research Council, the Vienna Science and Technology Fund, the European Space Agency, as well as by a number of industrial collaborations including with Intel Labs and InterDigital. He was the Chair of the Signal Processing for Communications and Networking Technical Committee of the IEEE Signal Processing Society in 2022, as well as of the UK \& Ireland Chapter of the IEEE Information Theory Society from 2017 to 2022. He was a Distinguished Lecturer of the IEEE Communications Society in 2021 and 2022, and he was a Distinguished Lecturer of the IEEE Information Theory Society in 2017 and 2018.   Prof. Simeone is the author of the textbook "Machine Learning for Engineers"  published by Cambridge University Press, four monographs, two edited books, and more than 200 research journal and magazine papers. He is a Fellow of the IET, EPSRC, and IEEE.  
\end{IEEEbiographynophoto}
\vfill
\end{document}